\documentclass{ws-mpla} 

\usepackage{epsfig}
\usepackage{color}

\begin{document}

\markboth{Akhilesh Ranjan, Hemwati Nandan} 
{}

\catchline{}{}{}{}{} 

\title{Modifications to the Hadronic Regge Trajectories} 
\author{\footnotesize Akhilesh Ranjan}
\address{Department of Physics, Manipal Institute of Technology, Manipal,
Udupi, 576104, Karnataka, India\\
akranjanji@rediffmail.com}

\author{Hemwati Nandan} 
\address{Department of Physics, Gurukula Kangri Vishwavidyalaya, Haridwar, 
249404, Uttarakhand, India\\
hntheory@yahoo.co.in}
\maketitle
\pub{Received (Day Month Year)}{Revised (Day Month Year)} 
\begin{abstract}
The effect of quark mass on the Regge trajectory is analysed. 
Modifications in the equations of Regge trajectories are shown for 
mesonic as well as baryonic systems. For mesonic systems, the Regge 
trajectories get modified, but still remain linear. Contrary to the 
mesonic case, the Regge trajectories for baryonic systems indicate 
non-linearity. It is shown that in low mass and angular momentum region  
two hadrons with different quark compositions can have same mass and 
angular momentum.
\end{abstract}
\ccode{PACS Nos.: 11.55Jy, 14.20.-c, 14.40.-n.} 
\section{Introduction}
All the known hadrons are composed 
of quarks with three different colors as indicated by the deep inelastic 
scattering experiments \cite{DIS}. The isolated colored quarks 
have not been observed in nature, i.e., they are confined to 
the interior of hadrons \cite{DIS,cheng} which leads to the well-known 
problem of quark confinement.  
The issue of confinement is still 
a very challenging problem to solve in theoretical physics. This is mainly 
because of the highly non-perturbative nature of strong interactions in low 
energy (infra-red) region \cite{cheng,bander,smilga,confinement95}. In this 
context, it is worth to notice the string model of confined quarks in hadrons 
resulting from the dual resonance model \cite{veneziano,giudice}. Such strings 
are characterised by the Regge trajectories of hadrons (i.e., the appropriate 
relationship between the classical mass $M$ and spin $J$ of hadrons). 
Similar picture of the string models from the view point of the linearly
rising Regge trajectories also emerges in dual superconductor models of 
QCD \cite{hemwati-1,hemwati-2,hemwati-3}.

The original motivation behind the Regge theory \cite{regge-1,regge-2} was 
to prove the validity of the Mandelstam representation 
\cite{mandelstam-1,mandelstam-2} in non-relativistic potential scattering. 
The Regge representation was assumed to hold in relativistic region
extended to the Mandelstam domain of analyticity for scattering 
amplitude in high energy hadron-hadron scattering. The moving
Regge poles were defined as `Regge trajectory' and soon a variety
of Regge trajectories were found to be nicely fitted to the observed
hadrons and resonances in terms of well known Chew-Frautschi
plots \cite{chew}. The quark bound states have linearly rising
Regge trajectories with numerous possible angular momentum states
for a given energy and offer crucial insight into the
mechanism of the quark confinement \cite{cheng,confinement95,perkins}. 

\begin{figure}[ph]
\centerline{\psfig{file=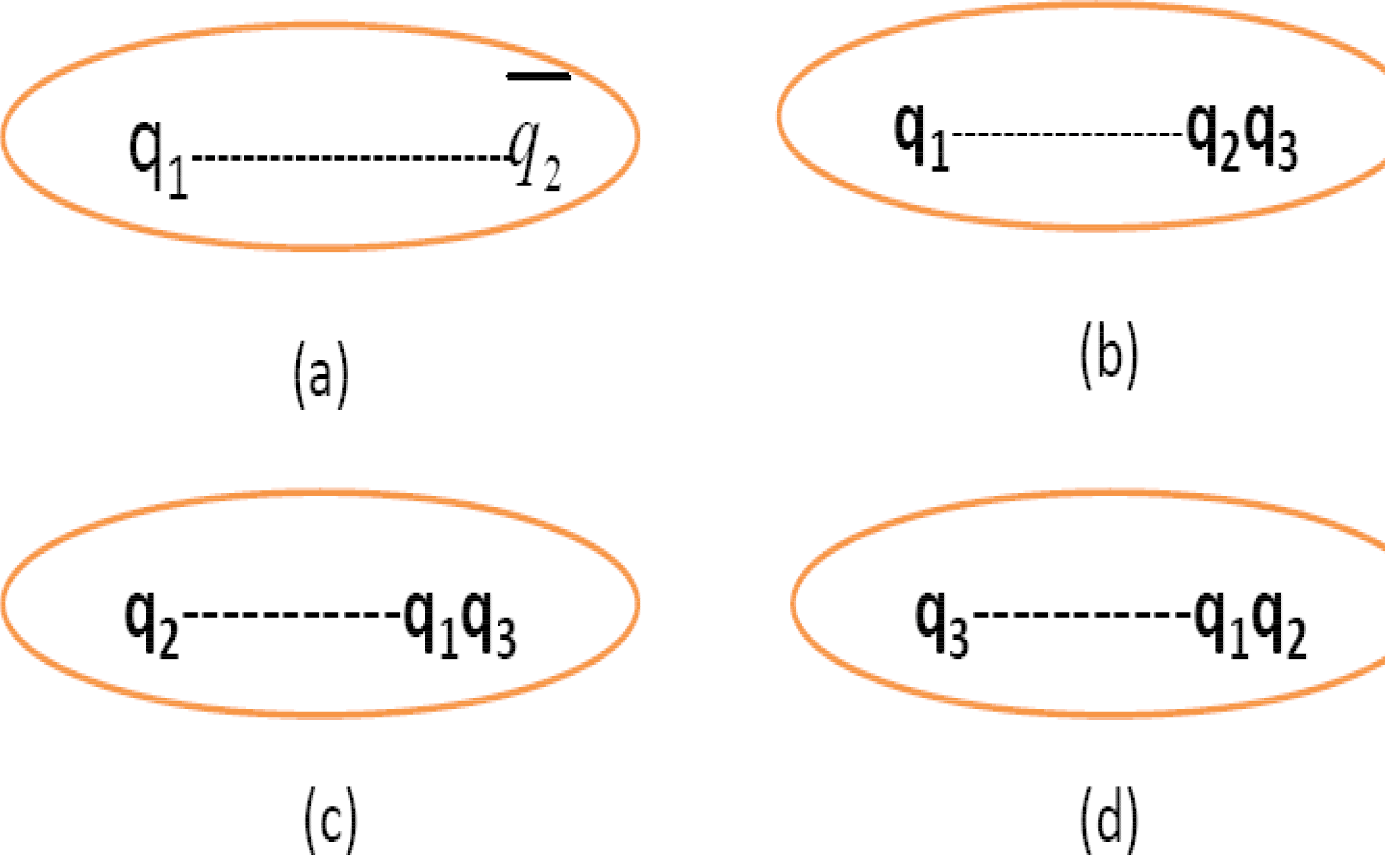,height=4.0 cm, width=8.0 cm}}
\vspace*{-2mm}
\caption{(a) String model representation of meson; (b),(c) \& (d) different 
possible configurations of a baryon in string model.\protect\label{hadron}} 
\end{figure} 

The observed relation between the angular momentum quantum number and 
the energy of a particular hadronic state is given by, 
\begin{eqnarray}
\label{JM}
J=\alpha' M^2+\alpha_{0} 
\end{eqnarray}
where, $\alpha_{0}$ is a constant known as the Regge intercept and 
$\alpha'$ is the Regge slope parameter which is given by 
$\alpha'={1}/({2\pi K})$ here, $K$ is the linear energy density of the 
string. The Eq(\ref{JM}) holds for the case of constant energy density 
$K$ of the string \cite{perkins}, that is, to a linear potential of 
the form $V(r)=Kr$ where, $r$ is the inter-quark distance. The linearly 
rising Regge trajectories are, therefore, relevant to the linear form of 
QCD potential and again confirm the validity of the string model of
hadrons. The observed value of Regge slope parameter $(\alpha'=0.93 GeV^{-2})$ 
then leads $K=0.87 GeV/fm$ and the same result also comes from the 
consideration of the measured size of hadrons in electron scattering 
experiments \cite{bethke}. 

  The microscopic understanding of the generic properties of the Regge 
trajectories for hadrons are still not completely explained. The study 
of the Regge trajectories \cite{regge-1,regge-2,inopin-1} may, therefore, 
be a crucial test for the physical realisation of the string model. The 
total angular momentum and classical mass of the strings for mesons and 
baryons (or hadrons) need careful attention especially in view of the 
universality of the Regge slope parameter of hadronic trajectories 
\cite{cheng}.  Besides the string model, there exists quark-diquark 
model for baryons \cite{diquark-1,diquark-2,diquark-3,diquark-4,diquark-5} and it would also be 
meaningful to work out the structure 
of Regge trajectories in the framework of such model. 
In recent years, many attempts have been made to address the 
problem specially the emergence of non-linearity in the Regge trajectories 
with their physical realisation in different models
\cite{inopin-1,hothi,bura,soloviev,inopin-2,sharov-1,sharov-2,nguyen}. 

  In this letter, we have analysed a more realistic string model of hadrons 
by considering the modifications in Regge trajectories due to the finite 
quark mass. It is shown that the linearity of the Regge trajectories remain 
intact in case of mesons while the Regge trajectories for baryons deviate 
from linearity by deriving the expressions for the classical mass and
classical angular momentum in both cases. 
\section{Calculations}
In string model of hadrons, the quarks are generally assumed to be massless 
and spinless. In usual Regge trajectories the quark-antiquark which are 
connected by a string of length  $l$, assumed to be rotating with the speed 
of light. In such case, the relation between the classical mass and classical 
angular momentum of a hadron is given by Eq(\ref{JM}). 

 We propose the following modifications for mesonic as well as 
baryonic Regge trajectories separately in view of the string model 
configurations presented in  Fig\ref{hadron}.
\subsection{Corrections for Mesonic Regge trajectories}  
Let us consider, a 
general case where a meson is made of two different flavors of  
quark and antiquark with finite mass. We assume that 
the quark and antiquark are sitting at the opposite ends of the string. 
Further, we do not consider the spin of the quarks. The mesonic string is 
rotating about its center of mass (see Fig 1(a)). Let the quark with mass 
$m_1$ is rotating with speed $fc$ where $0<f\leq1$ and $c$ is the speed of 
light in vacuum. Throughout our calculations, we have considered  
the natural units, i.e., $c=1$ or, $fc=f$. The mass of antiquark is $m_2$ 
and $l$ is length of the string. The modified mass of the meson ($M_{mod}$) 
is then given by 
\begin{eqnarray} 
\label{mes-m1}
M_{mod}=\frac{Km_2l}{f(m_1+m_2)}\left(\int_{0}^{f}\frac{\,dv}{\sqrt{1-v^2}}+\int_{0}^{\frac{m_1}{m_2}f}\frac{\,dv}{\sqrt{1-v^2}} \right)+\gamma_1m_1+\gamma_2m_2
\end{eqnarray} 
where $\gamma_1=\frac{1}{\sqrt{1-f^2}}$ and
$\gamma_2=\frac{1}{\sqrt{1-\frac{m^2_1f^2}{m^2_2}}}$. After integration 
Eq(\ref{mes-m1}), can be rewritten as
\begin{eqnarray} 
\label{mes-m2}
M_{mod}=\frac{Km_2l}{f(m_1+m_2)}\left(sin^{-1}f+sin^{-1}\frac{m_1f}{m_2}\right)+\gamma_1m_1+\gamma_2m_2  
\end{eqnarray} 
The modified angular momentum of meson ($J_{mod}$) is given by 
\begin{eqnarray}
\label{mes-ang-mom1} 
J_{mod}&=&\frac{Km_2^2l^2}{f^2(m_1+m_2)^2}\left(\int_{0}^{f}\frac{v^2\,dv}{\sqrt{1-v^2}}+\int_{0}^{\frac{m_1}{m_2}f}\frac{\,dv}{\sqrt{1-v^2}} \right)\nonumber\\
&+&\frac{m_1fl}{m_1+m_2}(\gamma_1m_2+\gamma_2m_1)  
\end{eqnarray} 
and after integration which leads to the following form 
\begin{eqnarray} 
\label{mes-ang-mom2}
J_{mod}&=&\frac{Km_2^2l^2}{f^2(m_1+m_2)^2}\left(\frac{1}{2}sin^{-1}f-\frac{f}{2}\sqrt{1-f^2}+\frac{1}{2}sin^{-1}\frac{m_1f}{m_2}\right.\nonumber\\ 
&-&\left.\frac{m_1f}{2m_2}\sqrt{1-\frac{f^2m_1^2}{m_2^2}}\right)
+\frac{m_1fl}{m_1+m_2}(\gamma_1m_2+\gamma_2m_1)
\end{eqnarray} 
Now the modified expression of the inter-relationship between the classical 
mass and classical angular momentum can be expressed as 
\begin{eqnarray}
\label{reg-tra3} 
J_{mod}&=&\frac{\pi}{sin^{-1}f+sin^{-1}\frac{m_1f}{m_2}}\left(1-\frac{sin(2sin^{-1}f)+sin\left(2sin^{-1}\frac{m_1f}{m_2}\right)}{2\left(sin^{-1}f+sin^{-1}\frac{m_1f}{m_2}\right)}\right)\nonumber\\ 
&\times&\alpha^{\prime}(M-\gamma_1m_1-\gamma_2m_2)^2\nonumber\\ 
&+&\frac{2\pi\left(\frac{m_1}{m_2}\right)f^2\alpha'}{sin^{-1}f+sin^{-1}\frac{m_1f}{m_2}}(M-\gamma_1m_1-\gamma_2m_2)(\gamma_1m_2+\gamma_2m_1)
\end{eqnarray}  
\noindent 
The expressions of $J_{mod}$ are functions of $sin^{-1}\frac{m_1f}{m_2}$. 
Since $sin\theta\le 1$ so $f\le\frac{m_2}{m_1}$. Further, according to 
special theory of relativity $f\le1$. Both of these conditions should 
therefore satisfy simultaneously. 
\begin{figure}[ph]
\centerline{\psfig{file=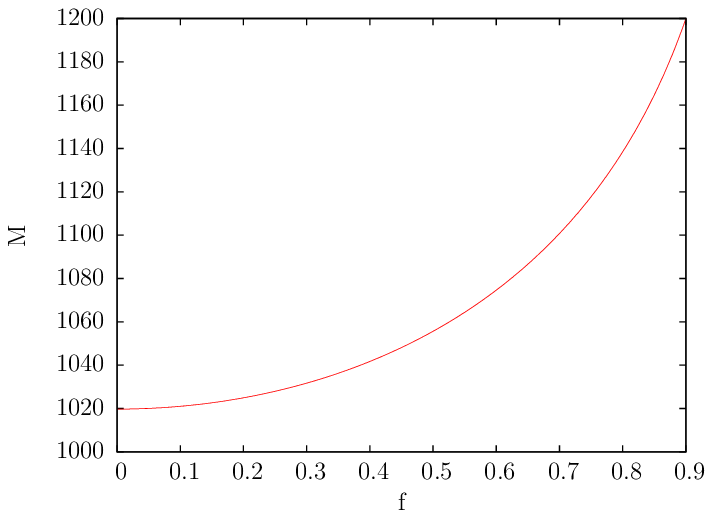,height=20.0 cm, width=15.0 cm}}
\vspace*{-72mm}
\caption{Mass of meson as a function of quark `i''s speed. Mass of 
quark `i' remains fixed.\protect\label{f-mfM}}
\end{figure} 

\begin{figure}[ph]
\centerline{\psfig{file=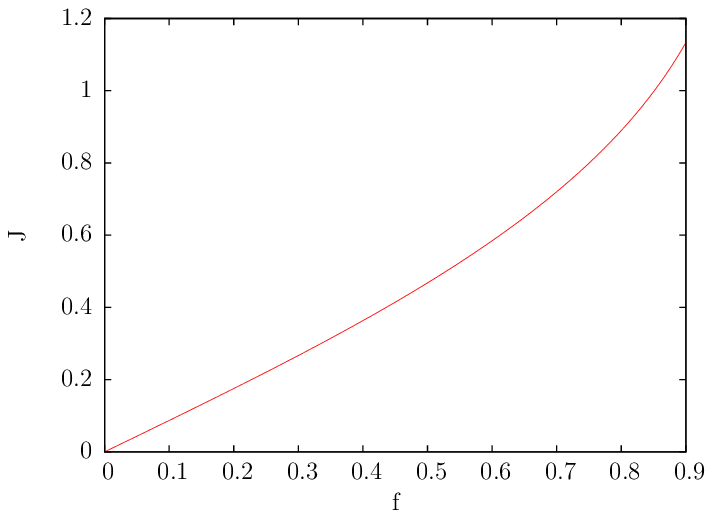,height=20.0 cm, width=15.0 cm}}
\vspace*{-72mm}
\caption{Angular momentum of meson as a function of quark `i''s speed. 
Mass of quark `i' remains fixed.\protect\label{f-mfJ}}
\end{figure} 

\begin{figure}[ph]
\centerline{\psfig{file=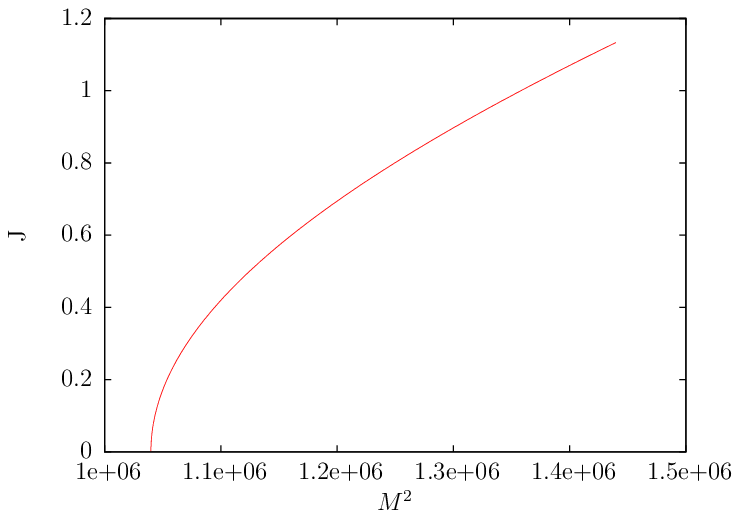,height=20.0 cm, width=15.0 cm}}
\vspace*{-72mm}
\caption{The Regge trajectory for mesons when the quark `i''s speed is 
varying. Mass of quark `i' remains fixed.\protect\label{f-m-MJ}} 
\end{figure} 

\begin{figure}[ph]
\centerline{\psfig{file=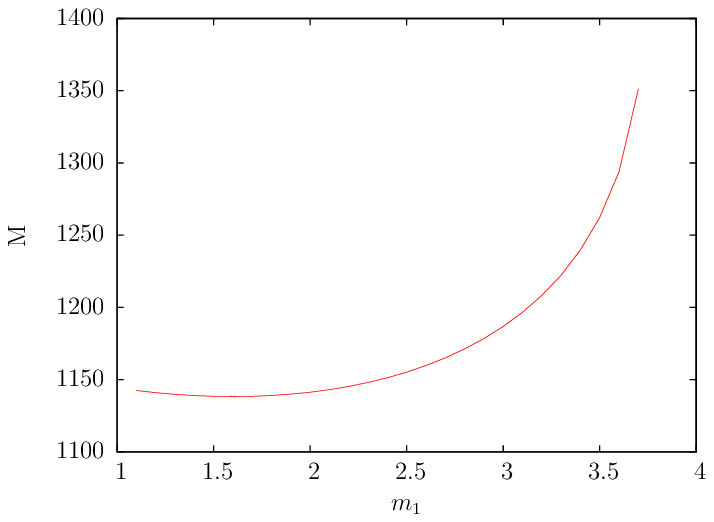,height=20.0 cm, width=15.0 cm}}
\vspace*{-72mm}
\caption{Mass of meson as a function of quark `i''s mass. Speed of
quark `i' remains fixed.\protect\label{m-mmM}} 
\end{figure} 

\begin{figure}[ph]
\centerline{\psfig{file=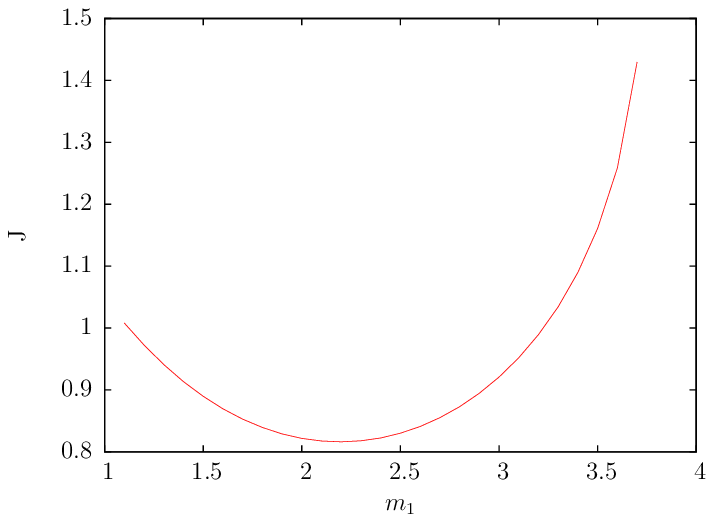,height=20.0 cm, width=15.0 cm}}
\vspace*{-72mm}
\caption{Angular momentum of meson as a function of quark `i''s mass.
Speed of quark `i' remains fixed.\protect\label{m-mmJ}} 
\end{figure} 

\begin{figure}[ph]
\centerline{\psfig{file=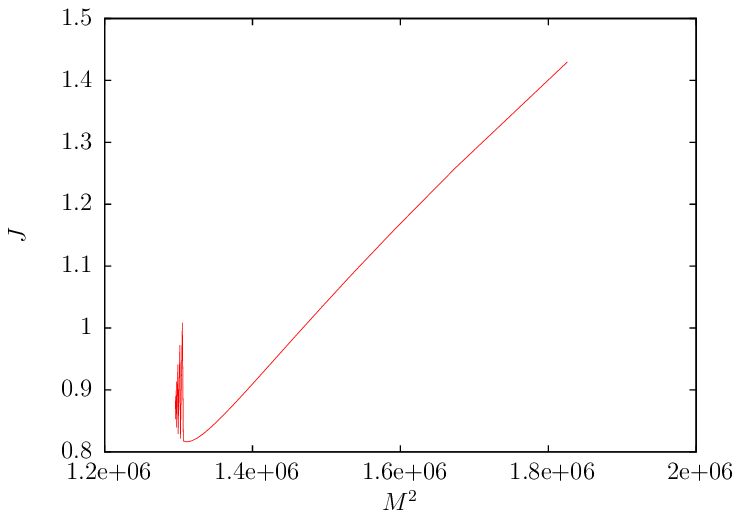,height=20.0 cm, width=15.0 cm}}
\vspace*{-72mm}
\caption{The Regge trajectory for meson when the mass of quark `i' is varying. 
Speed of quark `i remains fixed.\protect\label{cm-m-MJ}}  
\end{figure} 

\subsection{Correction for baryonic Regge trajectories} 
Like a string of meson, a baryon can also be described in terms of three 
equally probable configurations as shown in Fig 1.(b)-(d). On following the 
analysis of mesons, we obtain the modified expression for classical mass of 
a baryon for configuration (a) 
\begin{eqnarray} 
\label{bar-m1}
M_1=\frac{Kl(m_2+m_3)}{f(m_1+m_2+m_3)}\left(sin^{-1}f+sin^{-1}\frac{m_1f}{m_2+m_3}\right)+\gamma_0m_1+\gamma_a(m_2+m_3)   
\end{eqnarray} 
where $m_1, m_2, m_3$ are masses of quarks `1', `2', and `3' respectively and 
$\gamma_0=\frac{1}{\sqrt{1-f^2}}, \& \gamma_a=\frac{1}{\sqrt{1-\left(\frac{m_1f}{m_2+m_3}\right)^2}}$.\\ 
\noindent
All of the above configurations for a string of baryon are equally probable. 
Therefore the actual modified expression for mass of a baryon should be their  
average. 
 \begin{eqnarray} 
\label{bar-m5}
or,~~~M_{mod}&=&\frac{2Kl}{3f}sin^{-1}f+\frac{Kl}{3f(m_1+m_2+m_3)}\left\{(m_2+m_3)sin^{-1}\frac{m_1f}{m_2+m_3}\right.\nonumber\\ 
&+&\left.(m_1+m_3)sin^{-1}\frac{m_2f}{m_1+m_3}+(m_1+m_2)sin^{-1}\frac{m_3f}{m_1+m_2}\right\}\nonumber\\
&+&\gamma_0(m_1+m_2+m_3)+\gamma_a(m_2+m_3)+\gamma_b(m_1+m_3)\nonumber\\
&+&\gamma_c(m_1+m_2)
\end{eqnarray}
where we have assumed that `f' and `l' are same in all probable configurations. 
It is assumed that in configuration (a) the quark `1' rotates with speed `f', 
in configuration (b) the quark `2' rotates with speed `f', in configuration (c) 
the quark `3' rotates with speed `f', $\gamma_b=\frac{1}{\sqrt{1-\left(\frac{m_2f}{m_1+m_3}\right)^2}}$, and $\gamma_c=\frac{1}{\sqrt{1-\left(\frac{m_3f}{m_1+m_2}\right)^2}}$.\\ 

Similarly the modified expression of angular momentum for configuration 
(a) is obtained as 
\begin{eqnarray}
\label{bar-j1} 
J_1&=&\frac{\pi}{sin^{-1}f+sin^{-1}\frac{m_1f}{m_2+m_3}}\left(1-\frac{sin(2sin^{-1}f)+sin\left(2sin^{-1}\frac{m_1f}{m_2+m_3}\right)}{2\left(sin^{-1}f+sin^{-1}\frac{m_1f}{m_2+m_3}\right)}\right)\nonumber\\
&\times&\alpha^{\prime}\{M_1-\gamma_0m_1-\gamma_a(m_2+m_3)\}^2\nonumber\\
&+&\frac{m_1fl}{m_1+m_2+m_3}\{\gamma_am_1+\gamma_0(m_2+m_3)\}  
\end{eqnarray}  
Again since all the configurations are equally probable therefore the actual 
modified expression for angular momentum associated with the baryon should be 
their average. The general structure of $J_i$'s (where i=1,2,3 stands for 
configurations a,b, and c respectively) are given by 
\begin{eqnarray} 
\label{bar-j5} 
J_i=C_i^2(M_i-a)^2+\gamma m_ifl
\end{eqnarray} 
where $C_i$'s are functions of `$f$', `$l$', and $m_1,m_2,m_3$. 
Now it can be proved that it is not possible to write the expression for 
$J_{mod}$ in the form of Eq\ref{JM} for the case of baryons with finite 
quark masses.   

In the expressions of $J_i$'s, we encounter terms 
$sin^{-1}\frac{m_if}{\sum_jm_j-m_i}$ where i,j can take values 1,2,3. 
Since $sin\theta\leq1$ therefore leads to the condition 
$f\leq\frac{\sum_jm_j}{m_i}-1$ which alongwith $f\leq1$ should satisfy 
simultaneously. 

It is clear that if one tries to write the equation of the Regge trajectory 
for a mesonic system in the form of Eq\ref{JM}, the expressions of $\alpha'$ 
and $\alpha_O$ will also get modified and they will become functions of quark 
masses as well as their rotational speed. 
\begin{figure}[ph] 
\centerline{\psfig{file=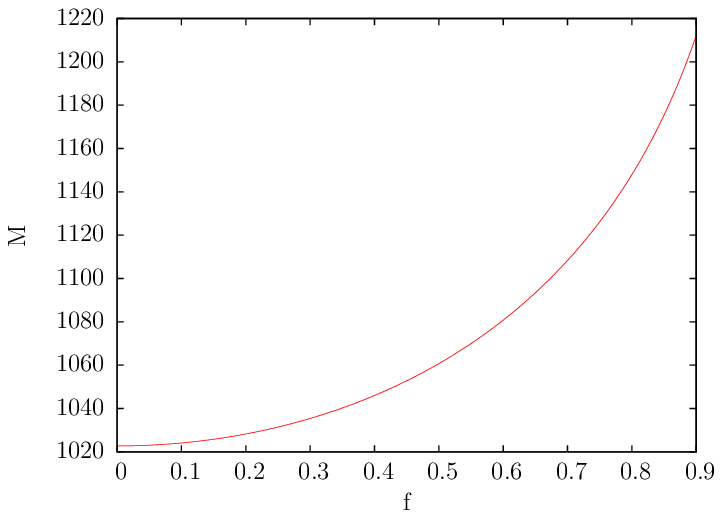,height=20.0 cm, width=15.0 cm}}
\vspace*{-72mm}
\caption{Mass of baryon as a function of quark `1''s speed. Mass of quark `1' 
remains fixed.\protect\label{f-bfM}} 
\end{figure} 

\begin{figure}[ph]
\centerline{\psfig{file=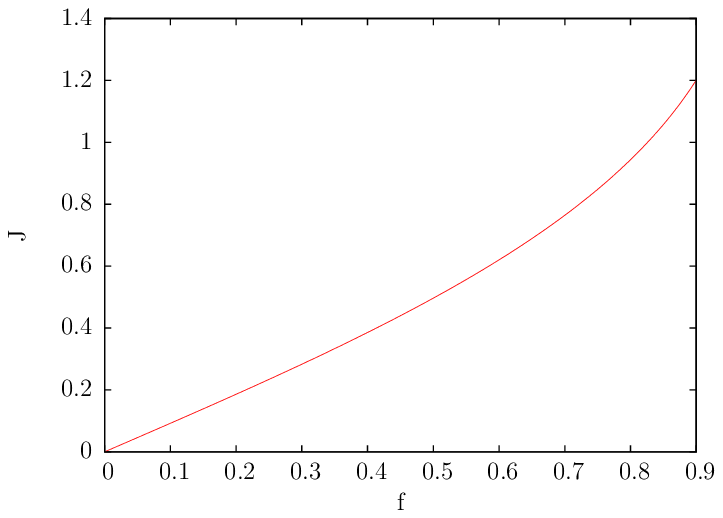,height=20.0 cm, width=15.0 cm}}
\vspace*{-72mm}
\caption{Angular momentum of baryon as a function of quark `1''s speed. 
Mass of quark `1' remains fixed.\protect\label{f-bfJ}}  
\end{figure} 

\begin{figure}[ph]
\centerline{\psfig{file=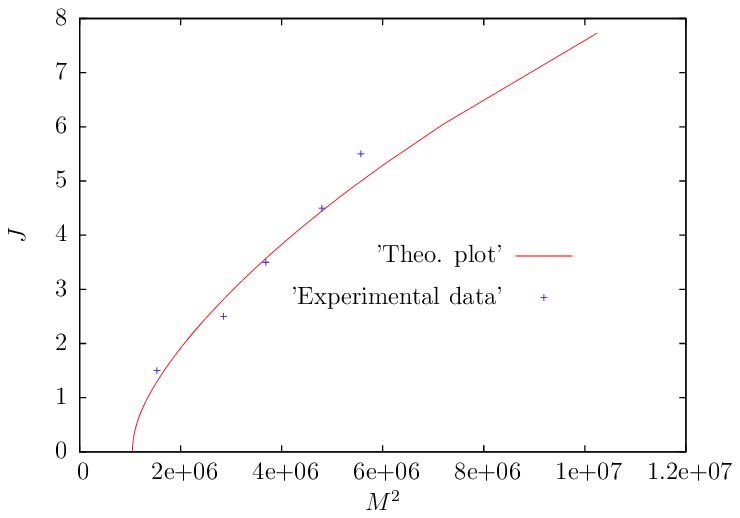,height=20.0 cm, width=15.0 cm}}
\vspace*{-72mm}
\caption{The Regge trajectory for baryons when the quark `1''s speed is 
varying. Mass of quark `1' remains fixed.\protect\label{cf-b-MJ}} 
\end{figure} 

\begin{figure}[ph]
\centerline{\psfig{file=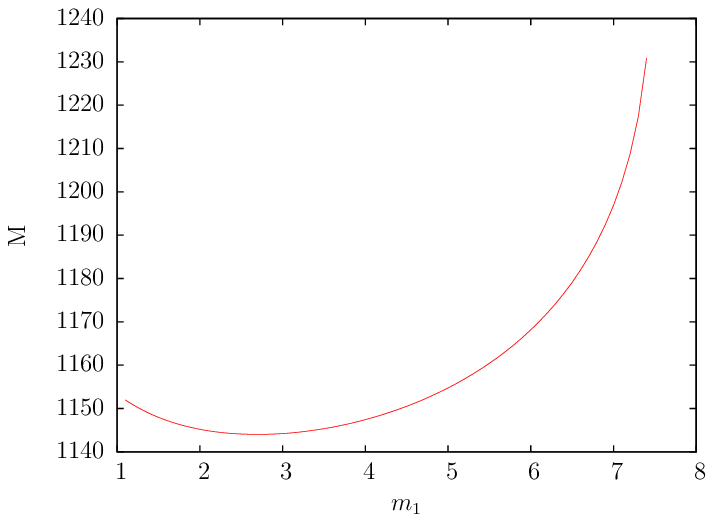,height=20.0 cm, width=15.0 cm}}
\vspace*{-72mm}
\caption{Mass of baryon as a function of quark `1''s mass. Speed of quark `1' 
remains fixed.\protect\label{m-bmM}}
\end{figure} 

\begin{figure}[ph]
\centerline{\psfig{file=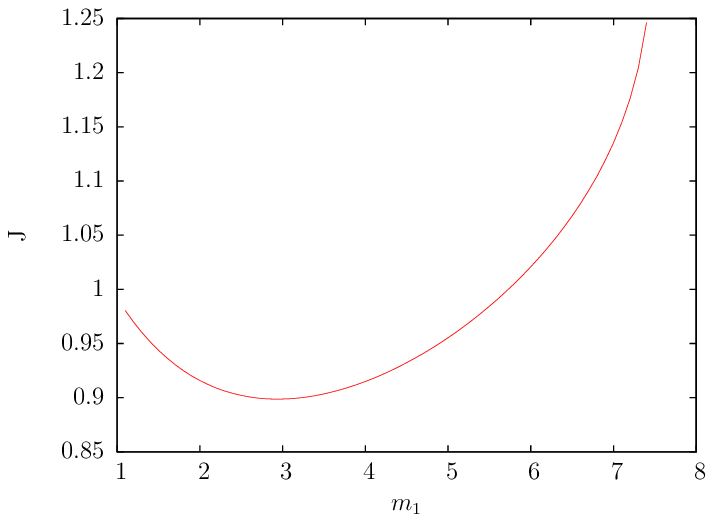,height=20.0 cm, width=15.0 cm}}
\vspace*{-72mm}
\caption{Angular momentum of baryon as a function of quark `1''s mass. Speed 
of quark `1' remains fixed.\protect\label{m-bmJ}}  
\end{figure} 

\begin{figure}[ph]
\centerline{\psfig{file=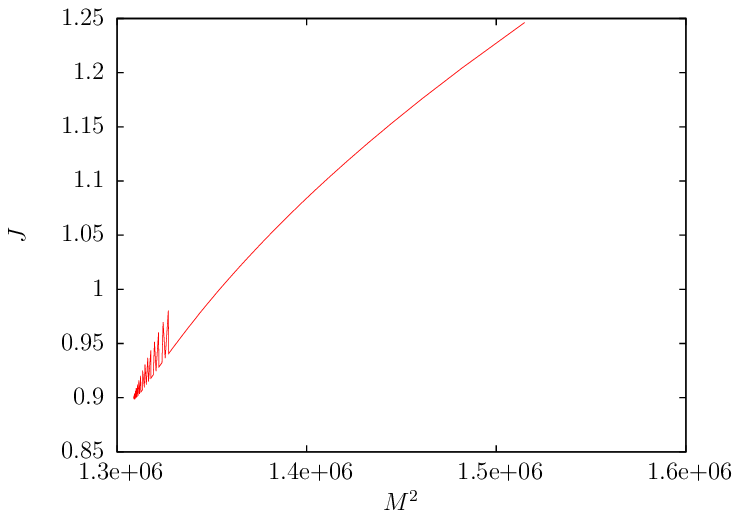,height=20.0 cm, width=15.0 cm}}
\vspace*{-72mm}
\caption{The Regge trajectory for baryon when the mass of quark `1' is 
varying. Speed of quark `1' remains fixed.\protect\label{cm-b-MJ}} 
\end{figure}
 
\section{Results and discussion}
The expressions obtained for the  inter-relationship between the classical 
mass and angular momentum for hadrons in the last section are visually 
presented in the Fig2 to Fig13 with different conditions 
on the mass of quarks and their rotational speed. 
In computations, we have considered $m_1=1.5MeV(u~quark), m_2=3.0MeV(d~quark), 
m_3=3.0MeV(d~quark), K=0.2GeV^2$, from Particle Data Group (PDG) 
\cite{quark-mass1}. For computation of mass of hadrons 
the string length $l=1fm$ is taken. In the plots mass of hadrons and 
corresponding angular momenta are calculated for different values of  
$f$ and quark mass $m_1$. Then Regge trajectories are plotted. 

In Fig\ref{f-mfM} and Fig\ref{f-mfJ} the dependence of meson mass and 
angular momentum on rotational speed of a quark are shown. Fig\ref{f-m-MJ} 
shows the mesonic Regge trajectory. 
All these plots show non-linear behavior quite well. It is worth noting that 
in Fig\ref{f-m-MJ} and Fig\ref{cf-b-MJ}, the plots are parabolic. In usual 
Regge trajectory with the assumption that the quarks are massless, the 
inter-relationship between $J$ and $M$  is given by equation \ref{JM} such 
that the plot $J$ vs $M^2$ is a straight line while in 
the present case with the consideration of massive quarks, the
inter-relationship between $J$ and $M$ acquires the form 
$J=\alpha (M-a)^2+(M-b)$.  The plot $J$ vs $(M-a)^2$ will be a straight line 
but since we have plotted $J$ vs $M^2$ therefore we obtain a parabolic 
trajectory. Hence in these figures, the non-linearity is not definitely
an artifact of the scale used. 

In Fig\ref{m-mmM} and Fig\ref{m-mmJ}, the dependence of meson mass and 
angular momentum on quark mass are shown. In these figures, the mass and 
angular momentum of meson first decreases with quark mass and after acquiring 
a minimum value it again starts increasing. In other words, these mesons with 
less mass and angular momentum can be described by two different kind of 
constituent quark-antiquarks. This leads to a confusing behavior of the Regge 
trajectory of less massive mesons where one mesonic mass can have two 
different angular momenta. But one should note that these two angular momenta 
are for two different mesons having same mass as explained earlier. 
From the Eq(\ref{reg-tra3}), it is clear that the Regge 
trajectory is linear, i.e., if we plot $J$ vs $(M-a)^2$ as explained above 
while the non-linear behavior of Regge trajectories is evident 
Fig\ref{cm-m-MJ}.  Since the linearly rising Regge trajectories 
are relevant to the linear form of confinement potential, it therefore 
remains to see the effect of the above mentioned non-linearity in Regge 
trajectories on the confinement potential at large distances in hadronic 
sense to check the validity of the string model of hadrons.

Let us compare Fig\ref{f-mfM} with Fig\ref{f-bfM}; Fig\ref{f-mfJ}
with Fig\ref{f-bfJ} and so on. One may notice that there are similarities 
and it is because for meson and baryon, we have analysed the similar kind of 
structures (i.e., string model) as shown in Fig\ref{hadron}. Therefore their 
physical interpretaion should remain same in both the cases. Hence in the 
Fig8 to Fig13 which describe the string model results 
and the Regge trajectories for baryons, the same pattern like of mesons 
is evident. 
For instance, we have also compared the nucleon resonance data 
$N^*(1238, \frac{3}{2}^+)$, $N^*(1688, \frac{5}{2}^+)$, 
$N^*(1920, \frac{7}{2})$, $N^*(2190, \frac{9}{2})$,  and 
$N^*(2360, \frac{11}{2})$ \cite{perkins,quark-mass1,quark-mass2} with our 
calculations corresponding to baryonic Regge trajectories and found them in 
close agreement as shown in Fig\ref{cf-b-MJ}. 
\section{Conclusions and future scope}
In this article, our primary purpose has been to see the effect 
of quark masses on Regge trajectories of hadrons. We have also shown, in the 
cases of mesons and baryons, how the Regge trajectories change with the 
variation of the mass of quarks as well as their rotational speed.
The important conclusions drawn from this study are summarised as follows. 
The consideration of the  massive quarks in the string model of hadrons, the 
expressions for the classical mass and classical angular momentum and hence 
the Regge trajectories are modified. For mesons, the form of the Regge 
trajectory (i.e., their linear behavior) remains same, but for the baryons 
the linearity is altered significantly. For mesonic systems, they still show 
linear nature but for baryonic systems they have highly non-linear behavior 
as presented in the corresponding figures. The parameters to characterise the 
Regge trajectory, i.e., $\alpha_o$ and $\alpha^{\prime}$ are also modified 
and become the quark mass dependent quantities. In low mass and angular 
momentum region two hadrons with different quark compositions can have same 
mass and angular momentum. The Regge trajectories for the present 
model are also compared with the trajectories corresponding to measured baryon 
masses in terms of the nucleon resonance data and are found in close 
agreement. 

The deviation from the known experimental results may be due to exclusion 
of spin of quarks. Since the baryons are three body systems and the exact 
analysis for a three body system is an extremely complicated problem in 
itself. It might be possible that the non-linearity in baryonic Regge 
trajectories have a close connection with the mass difference between the 
quark and diquark masses in baryons in comparison to the linear mesonic 
trajectories  with the equality of the quark and antiquark mass in mesons. 
This issue needs a careful attention. One so need to work out whether 
the non-linearity of Regge trajectories is dependent on the position-dependent 
quark masses. The analysis of modified Regge trajectories is an important task 
to study in greater detail especially from the view point of the data available 
for multitude of hadrons with the particle data group (PDG). We intend to 
report on such issues in our forthcoming communications. 

\section{Acknowledgement}
AR is thankful to Gurukula Kangri Vishwavidyalaya Haridwar and Harish-Chandra 
Research Institute Allahabad for their hospitality where some part of this 
work was done. We are grateful to the people of India for their generous 
support for research. Authors are also thankful to the anonymous referees 
for their critical and valuable suggestions which helped in improving the 
presentation of the paper.

\end{document}